\documentclass[9pt,twocolumn,twoside]{optica}
\setboolean{shortarticle}{false}
\setboolean{minireview}{false}
\usepackage[font=scriptsize]{caption}
\usepackage{float}
\usepackage{siunitx}
\usepackage{todonotes}
\usepackage{xfrac}
\title{Enhanced deep detection of Raman scattered light by wavefront shaping}

\author[1]{Alba M. Paniagua-Diaz}
\author[1]{Adrian Ghita}
\author[1]{Tom Vettenburg}
\author[1]{Nick Stone}
\author[1,*]{Jacopo Bertolotti}
\affil[1]{University of Exeter, Stocker Road, Exeter EX4 4QL, United Kingdom}

\affil[*]{Corresponding author: j.bertolotti@exeter.ac.uk}

\ociscodes{290.0290 Scattering; 170.5660 Raman spectroscopy; 170.6510 Spectroscopy, tissue diagnostics}

\begin{abstract}
Light scattering limits the penetration depth of non-invasive Raman spectroscopy in biological media. While safe levels of irradiation may be adequate to analyze superficial tissue, scattering of the pump beam reduces the Raman signal to undetectable levels deeper within the tissue. Here we demonstrate how wavefront shaping techniques can significantly increase the Raman signal at depth, while keeping the total irradiance constant, thus increasing the amount of Raman signal available for detection.
\end{abstract}

\setboolean{displaycopyright}{true}

\begin{document}

\maketitle

\section{Introduction}
Raman spectroscopy enables label-free identification of chemical composition~\cite{raman1928new,graves1989practical}. This is particularly useful for investigating living biological tissues, where the unique rotational and vibrational modes of organic molecules allow non-invasive detection of physiologically important and disease specific bio-analytes~\cite{mahadevan1996raman}. However, since the spontaneous Raman signal is directly proportional to the pump intensity that decreases rapidly with depth in diffusive media due to multiple-scattering, this limits how far into the tissue a specific molecular target can be detected. Innovative illumination and detection geometries may improve the signal level~\cite{matousek2016development}; however, the amount of light that can reach the target remains a fundamental limit. This effectively confines Raman spectroscopy to surface studies of scattering tissue~\cite{matousek2009emerging, Ghita2016, Kong2015}.

Although multiple scattering of light in disordered media can appear random, the underlying physics is completely deterministic and relatively well understood~\cite{Akkermans2007}. In recent years, wavefront shaping techniques have emerged as a way to manipulate waves, even through strongly scattering media~\cite{Vellekoop2007FocusingMedia,mosk2012controlling}. In particular, wavefront shaping allows the focusing of light at a specific target inside a scattering medium~\cite{vellekoop2008demixing}. This makes it possible to concentrate the pump on the Raman active material, even when it is buried deep inside a tissue~\cite{Thompson2016}. The limit of this approach is that wavefront shaping in an unknown multiply scattering structure is typically achieved iteratively, using the signal as a feedback and gradually improving it~\cite{Vellekoop}. As the Raman signal is generally weak, an adequate signal-to-noise ratio requires long integration times. The necessity to repeat the measurement many times makes this approach slow and limits the maximum enhancement achievable. In particular the pioneering work by Thompson et al. achieved an increase of the detected Raman signal by approximately $25\%$, but required at least 10-20 times the time of a single Raman measurement~\cite{Thompson2016}. Additionally, these approaches are only useful when there is an initial detectable Raman signal, not addressing the problem of blind sensing in scattering media.

In this article we show that wavefront-shaping the pump beam to maximize its penetration depth can overcome the two main limitations of techniques based on light focusing: speed, and compatibility with blind sensing. Given that in this case the feedback is provided by the pump and not the Raman signal, there is no need for prior knowledge of Raman elements. At the same time, due to the much larger intensity of the pump with respect to the Raman signal, the integration time in the detector can be much shorter. This enables  full optimization on a much shorter timescale than the typical Raman measurement~\cite{conkey2012high}. We experimentally show an enhancement of approximately $50\%$ through a scattering layer with optical density of $\sim48$, and study how the achievable enhancement in depth depends on the target position and the wavefront shaping quality.

\section{Pump energy distribution and Raman signal generation}
The propagation of light intensity in a scattering medium can be described by the diffusion equation:
\begin{equation}
\frac{d I}{d t} = D \nabla^2 I + S ,
\label{eq:diffusion}
\end{equation}
where $S$ is the source term and $D$ is the diffusion coefficient~\cite{Akkermans2007}. For a source not varying in time and a slab geometry, the steady-state solution for a point source (or Green's function) is given by:
\begin{equation}
g(z,z_j)=\frac{I_0}{D}\left(\frac{(z + z_{e1})(L+z_{e2}-z_{j})}{L+z_{e2}+z_{e1}} + (z_{j}-z)H(z-z_{j})\right) , 
\label{eq:green}
\end{equation}
where $I_0$ is the total intensity of the pump, $L$ is the thickness of the slab, $z_j$ is the injection length, representing the position of the point source, typically on the order of the transport mean free path of the scattering media $\ell_t$, and $H$ is the Heaviside step function. In order to obtain Eq.~\ref{eq:green} it is also necessary to set the boundary conditions. By imposing the conservation of energy fluxes in the medium one obtains Robin boundary conditions, which can be recast as $I=0$ at a distance from the edge of the sample (the extrapolation lengths $z_{e1}$ and $z_{e2}$)~\cite{Zhu1991}.

\begin{figure}[tb]
\centering
\includegraphics[width=1\linewidth]{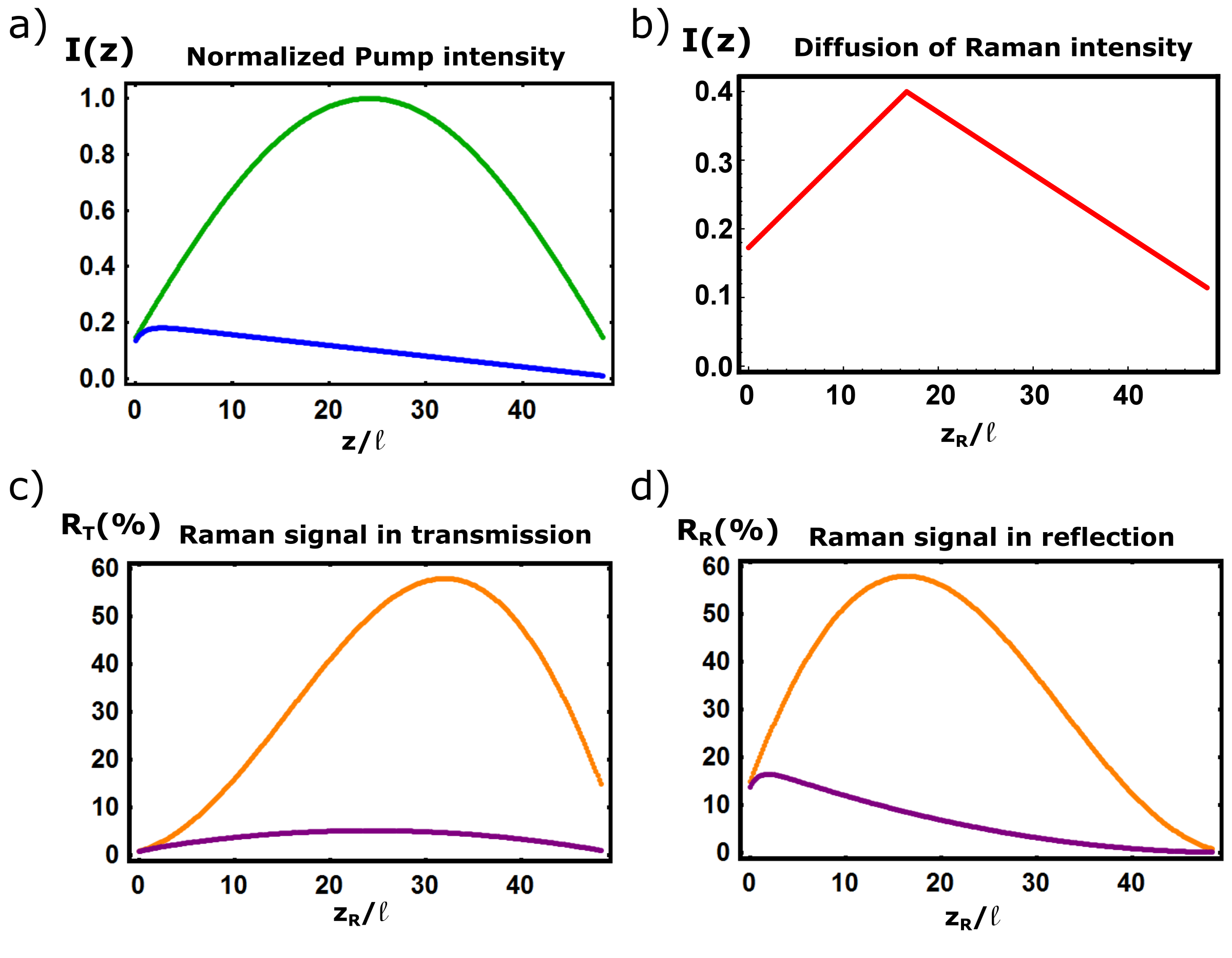}
\caption{a) Normalized intensity distribution of the pump light when the beam is not optimized (blue) and when all the incident energy is optimized and coupled to the fundamental solution of the diffusion equation (green). b) Intensity distribution of a point Raman element placed at position $z_R=10$. c) Raman scattered light in transmission of a target at position $z_R/\ell$, when the beam is optimized (orange) and when it is not optimized (purple). d) Raman scattered light in reflection of a target at position $z_R/\ell$, when the beam is optimized (orange) and when it is not optimized (purple). These results correspond to the solution to diffusion equation for a scattering medium with the same parameters than our experimental sample, described in Section~\ref{SectionExp}.}
\label{fig:RamanPump}
\end{figure}

The intensity profile $I(z)$ inside the diffusive slab is then given by the convolution of $g$ with the source $S$. There are two source functions that are relevant to the problem we are investigating: an exponentially decaying source $S(z)=e^{-z/z_{j}}$ due to the exponential decay of ballistic intensity as propagating through the medium (following Lambert-Beer law), and a point source $S(z)=\delta(z-z_{j})$, representing the diffusion process of an element originated at a specific position $z_j$. The intensity of the pump is well described by the exponentially decaying source (blue curve in Fig.~\ref{fig:RamanPump}a, whereas the intensity originated by a point Raman element, is better described by the point source (Fig.~\ref{fig:RamanPump}b). 

As energy conservation must hold under the assumption of non-absorbing media, a larger $T$ comes from a smaller fraction of reflected light $R$, which implies an increase in the intensity inside the medium, and therefore a larger pump penetration depth. This suggests that, if the total amount of transmitted light can be increased, the amount of pump light reaching the Raman active material embedded in the medium should increase proportionally, as would the generated Raman signal.

Interestingly, the diffusion approximation used to describe light propagation in scattering media does not take into account the fact that light is a wave. However, the multiply scattered light paths inside a disordered medium can interfere, giving rise to a large number of effects, from speckle correlations~\cite{berkovits94,deboer92}, the coherent backscattering cone~\cite{Albada1985,Wolf1985}, to the presence of open and closed channels~\cite{beenakker1997random}. This last effect is particularly important here, as it means that, with the right incident wavefront, it is possible to transmit all the incident light through a scattering medium~\cite{Dorokhov1984}. In contrast to what one might expect, maximizing the total transmission does not mean that the light traverses the sample unhindered. On the contrary, the mode with the highest transmission is also one that accumulates a large amount of energy inside the sample~\cite{choi2011transmission,Davy2015}, and the energy distribution takes an approximately cosine shape (green curve in Fig.~\ref{fig:RamanPump}a), with a maximum at the center~\cite{koirala2017inverse}.

As shown in Fig.~\ref{fig:RamanPump}a, the diffusion approximation and the open channels represent the worst and best case scenarios, respectively. The diffusion solution in a uniform slab (blue curve) has a maximum at roughly one transport mean free path of depth, but then decreases rapidly with depth, meaning that not much light is able to reach the deepest parts of the medium. In contrast, if one could perform perfect wavefront shaping and couple to an open channel (green curve), the pump distribution would be symmetric and have a maximum at the center. This is a particularly interesting feature, given that when the element to detect is in the central regions of the sample, the pump intensity is generally minimal, and consequently it becomes much more difficult to detect Raman signal from there~\cite{vardaki2015studying}. Realistic or experimental wavefront shaping will produce an intermediate distribution between the open channel and diffusive extremes~\cite{Davy2015}.

Once the pump reaches the Raman-active material, the Raman signal is generated and propagates diffusively to the exit of the sample. For simplicity we assume that the Raman signal is generated by a point source at position $z_R$, resulting in a distribution of the Raman signal given by Eq.~\ref{eq:green} with $z_j=z_R$ (Fig.~\ref{fig:RamanPump}b). 
The total amount of diffused light intensity that exits the medium is given by its gradient at the interface (Fick's law~\cite{fick1855}), so we can define $T$ and $R$ as the intensity coming out from the transmission and reflection side of the slab respectively (normalized to the incident intensity), as~\cite{Akkermans2007,Ojambati2016CouplingMedium}:
\begin{equation}
T=-\frac{D}{I_0}\left.\frac{d I(z)}{d z}\right|_{z=L} \quad \text{and} \quad R=\frac{D}{I_0}\left.\frac{d I(z)}{d z}\right|_{z=0} .
\label{eq:totransref}
\end{equation}

Using Eq.~\ref{eq:totransref} we can then calculate the total amount of diffuse Raman signal that can be collected in transmission ($R_T$) and in reflection ($R_R$) as a function of $z_R$ when the different pumps are considered (Fig.~\ref{fig:RamanPump}c-d). As seen in Fig.~\ref{fig:RamanPump}c-d, if the Raman active material is very close to the surface of the material there is no real advantage from the wavefront shaping approach. On the other hand, for depths larger than a few transport mean free paths, we expect a significantly higher signal if we wavefront-shape the pump light to maximize transmission, particularly for centered elements.

\section{Wavefront shaping of the pump for maximal transmission}
\label{SectionExp}
Despite the great interest and efforts put towards fully transmitting the incident light through scattering media (or coupling to their open channels), it has not been possible to experimentally achieve it so far, although substantial increase in the transmitted light has been reported~\cite{kim2012maximal, choi2015preferential, Ojambati2016CouplingMedium, Vellekoop2008UniversalMaterials}. The main factor preventing the experimental coupling of light to the open channels is the limited control of degrees of freedom of the medium~\cite{goetschy2013filtering}, which makes the singular values of the transmission matrix follow the Marchenko-Pastur distribution, typical of uncorrelated matrix elements. 

The total number of degrees of freedom at the entrance of the sample can be well approximated by the number of diffraction limited spots contained in the illumination area (A), such that: $N \propto \frac{2 \pi A}{\lambda^2}$, where $\lambda$ is the wavelength of the illumination source~\cite{Vellekoop2008UniversalMaterials}. In our experimental case, the radius of the illuminated spot at the entrance of the sample is $R=\SI{22}{\micro\meter}$ and $\lambda=\SI{785}{\nano\meter}$, which makes $N\approx 15500$. However, due to the limited degrees of control segments of the spatial light modulator, the number of controlled degrees of freedom at the entrance of the sample is $N_C = 3520$. Following from the Marchenko-Pastur distribution~\cite{goetschy2013filtering} the optimal enhancement in total transmission that can be achieved is given by~\cite{Marcenko1967}:
\begin{equation}
\left\langle \frac{T_{op}}{T_{in}}\right\rangle=\left(1+\sqrt{\frac{N_C}{M}}\right)^2
\label{totransm}
\end{equation}
where $M$ is the number of degrees of freedom at the output of the sample (in the output illuminated area). Assuming light diffusion at approximately $45^\circ$, the number of modes $M$ for our sample (see Section 4), is approximately $80000$ at the output of the sample. We can thus expect to increase the total transmission by a factor of $1.46$. 

\section{Materials and Methods}

\begin{figure}[tb]
\centering
\includegraphics[width=0.8\linewidth]{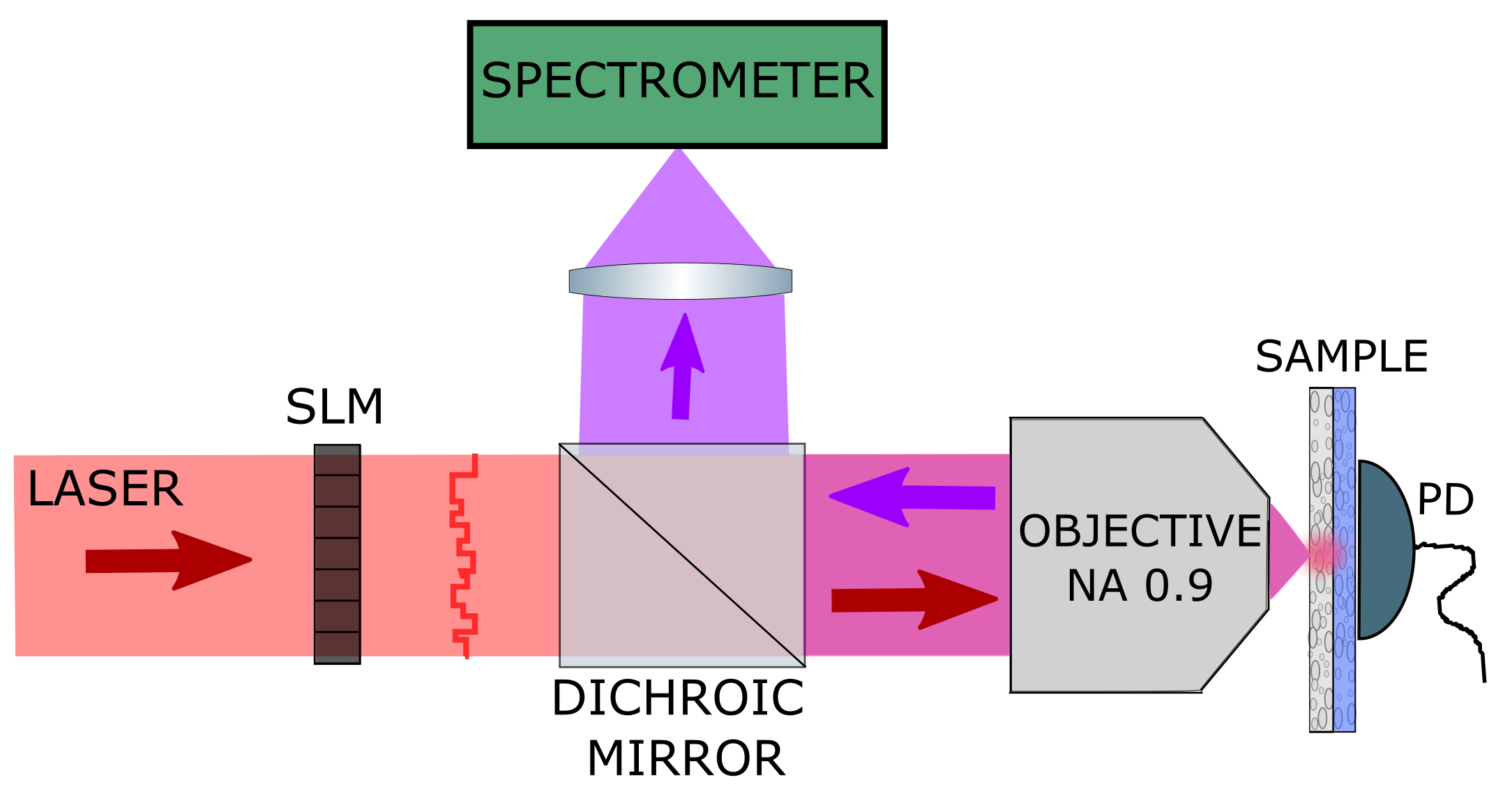}
\caption{Schematic of the experimental apparatus used to increase total transmission through a scattering medium whilst the Raman spectra is collected in reflection. Red arrows represent the pump light at $\SI{785}{\nano\meter}$ and the purple arrow represents the longer wavelengths generated due to the spontaneous Raman scattered light.}
\label{fig:expsetup}
\end{figure}

The experimental apparatus is shown in Fig.~\ref{fig:expsetup}. The pump laser ($\SI{1}{\watt}$ MOPA, Innovative Photonic Solutions) has its central wavelength in the near infrared $\lambda=\SI{785}{\nano\meter}$, to achieve larger penetration depth in biological tissue. The wavelength region that allows maximal penetration in the optical regime is roughly between \SI{700} and \SI{900}{\nano\meter}~\cite{Jacques2013,vo2014biomedical} where the absorption band of water has a broad dip. The pump laser is incident on a Digital Micromirror Device (DLP 9500, Texas Instruments) spatial light modulator (SLM) that iteratively adapts the phase profile of the wavefront. The modulated wavefront is focused on the sample by a microscope objective (NA=0.9, 100x, dry, Leica), achieving a spot size of $44\pm\SI{12}{\micro\meter}$ in diameter.  A photo-diode collects the total transmitted light (mainly from the pump) and utilizes it as feedback for an iterative algorithm controlling the SLM (see Supplement 2 for a detailed description of the algorithm). The backscattered Raman signal is filtered by a dichroic mirror (LPD02-785RU, Semrock), together with a long pass filter (LP02-785RU-25, Semrock). After the filtering phase, the Raman light is directed onto a spectrometer (Holospec 1.8i, Kaiser optical) equipped with a high dispersion grating ($\SI{864}{\milli\meter}$, Kaiser Optical) and a NIR CCD detector (Andor iDus 420, Oxford Instruments). The estimated spectral resolution of the system is $\SI{1.08}{\centi\meter^{-1}}$ with a $\SI{25}{\micro\meter}$ slit, measured with a calibration atomic line source.

\begin{figure}[tb]
\centering
\includegraphics[width=1\linewidth]{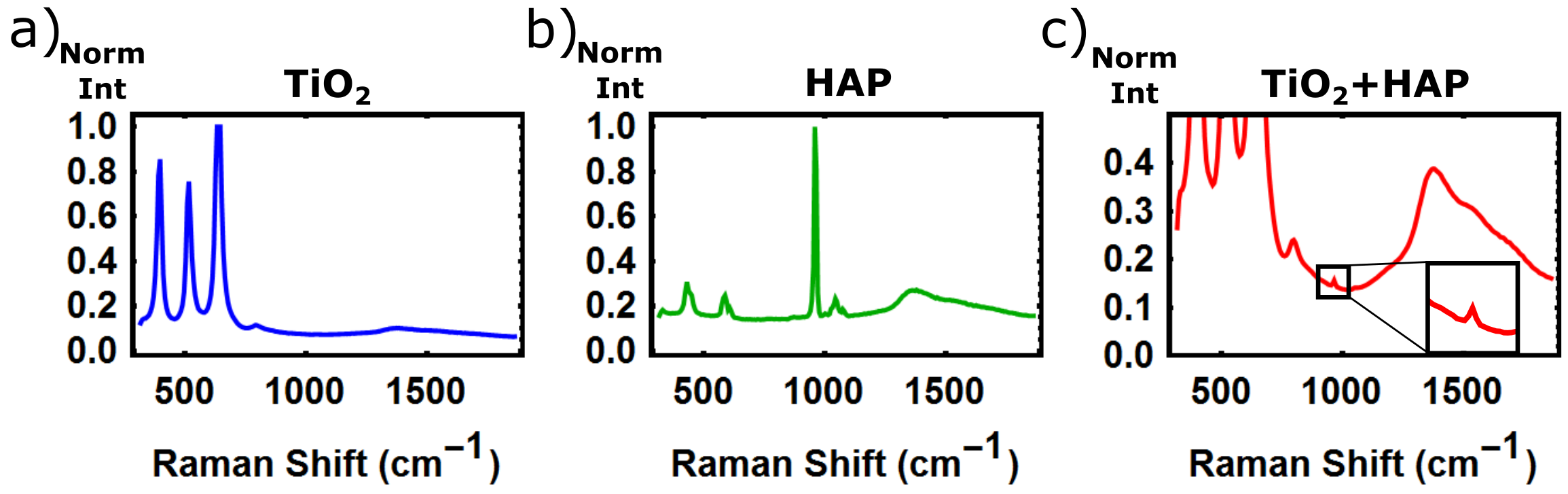}
\caption{Normalized Raman spectra of the different materials of the sample: a) Anatase TiO$_2$ spectrum, with three main peaks at $\SI{396}, \SI{512} and \SI{631}{\centi\meter^{-1}}$  b) Calcium Hydroxyapatite (HAP) with the main peak at $\SI{960}{\centi\meter^{-1}}$ and c) the collected spectrum of the combined sample of TiO$_2$ and HAP, where we can see a strong signal coming from the first layer of TiO$_2$ and a weak peak coming from the inner HAP at $\SI{960}{\centi\meter^{-1}}$, magnified in the inset. The broad contribution around $\SI{1500}{\centi\meter^{-1}}$ is due to the fluorescence of the microscope cover slide.}
\label{fig:combspec}
\end{figure}

The sample was made of two different materials, a highly scattering layer of anatase TiO$_2$ (Fisher Scientific UK) of thickness $L=\SI{29}{} \pm \SI{2}{\micro\meter}$ and a second layer of Calcium Hydroxyapatite (HAP) with a thickness of  $\SI{120}{} \pm \SI{10}{\micro\meter}$ (Sigma Aldrich UK). The layer of TiO$_2$ was chosen because it is a highly scattering material with very little absorption, which allows us obtain large optical densities with small thicknesses, in this case $OD=L/\ell_t\sim 29/0.6 \sim 48$, comparable to thicknesses of $\sim \SI{5}{\centi\meter}$~\cite{Jacques2013,vo2014biomedical} in fatty biological media, where the  transport mean free path is $\sim$ \SI{1}{\milli\meter} (the effect of a thicker sample is considered in the Discussion section). In this article we consider only non-absorbing scattering media; a generalization to absorbing media would modify Eq.~\ref{eq:green} introducing an exponential decay of the intensity from the source origin, and would change the shape of the intensity distribution of the open channels~\cite{Cao2015openchannelsabs}, but would not change the physics of the problem. HAP is a calcium phosphate mineral similar to that found in many hard tissues, and of interest for various biomedical applications~\cite{mucalo2015hydroxyapatite}, for instance as a marker to detect calcifications in breast tissue~\cite{matousek2009emerging}. Fig.~\ref{fig:combspec} shows the Raman spectra for the anatase TiO$_2$ and HAP materials (panels a and b), and panel c shows the combined spectra collected by the spectrometer in the experimental reflection configuration, where we can see the reduced peak from HAP at around $\SI{960}{\centi\meter^{-1}}$, attenuated by the TiO$_2$ layer.
The spectra were acquired in 5 accumulations of 10 seconds for each acquisition. The laser power incident on the sample was measured to be $\SI{9.7}{} \pm \SI{0.1}{\milli\watt}$. After removing the offset intensity from the detector, the recorded Raman spectra were imported to Mathematica for data post-processing, primarily consisting of background subtraction (third order polynomial). After background subtraction we performed a Lorentzian fit to the peak of interest and normalized the values for simplicity, as shown in Fig.~\ref{fig:Inc40}. In order to evaluate the increase in the Raman peak of interest, we compare the integrated area under the peak for the optimized and non-optimized wavefronts. 

\section{Results}

\begin{figure}[tb]
\centering
\includegraphics[width=1\linewidth]{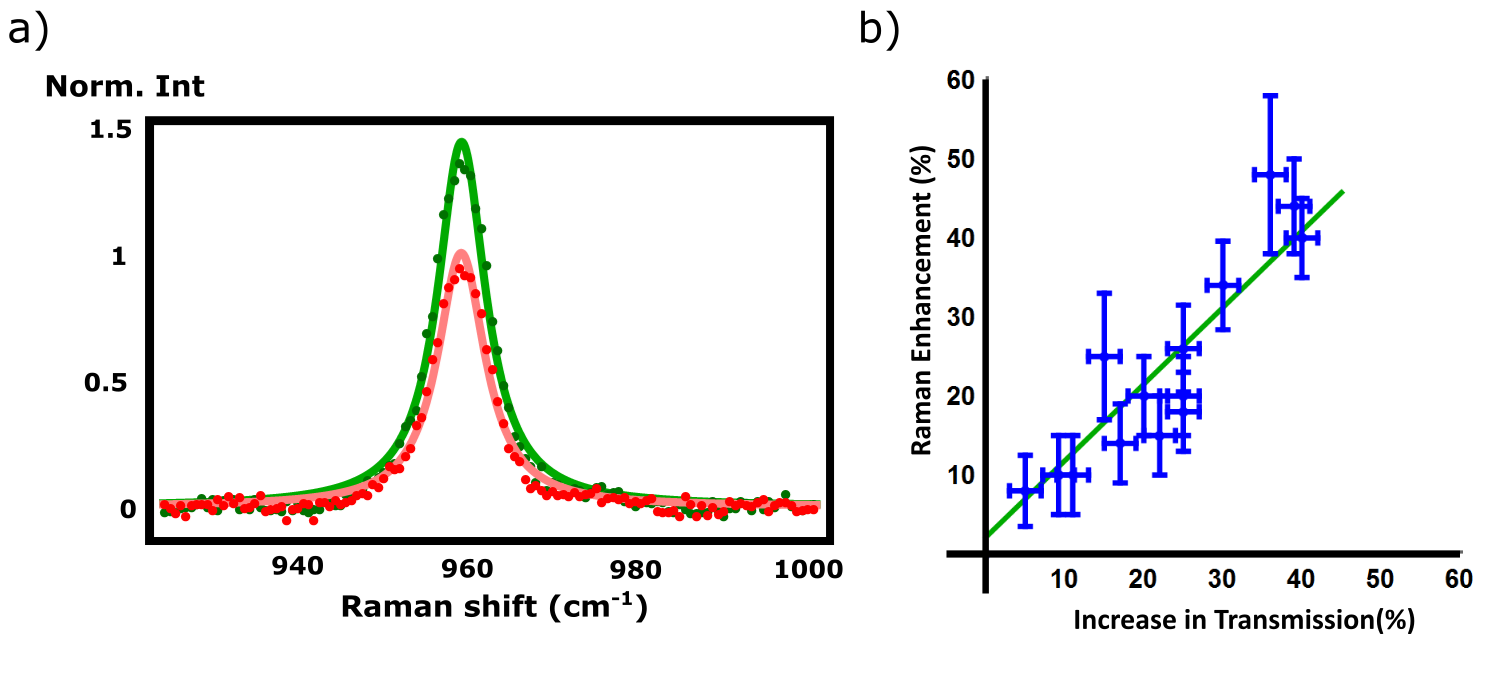}
\caption{a) Normalized spectral data points of the HAP peak collected in reflection before (red) and after (green) the wavefront optimization. Dots represent the experimental data and the solid lines, the Lorentzian fits to the data. The data in the figure corresponds to an enhancement in the Raman signal of 41\%, which was achieved with an increase in total transmission of 39\%. b) Increase of the Raman signal against the increase in the total pump transmission. It is possible to see how the data points follow a linear trend (in green) with a slope of 1.07 $\pm$ 0.12.
}
\label{fig:Inc40}
\end{figure}

Using iterative wavefront shaping techniques we increased the total transmission of the pump light through the scattering medium by a factor $1.40 \pm 0.3$. In Fig.~\ref{fig:Inc40}a we show the Raman signal intensity corresponding to the main peak of the subsurface HAP layer, before (red) and after the wavefront optimization (green). The dots represent the experimental data and the solid line curves are Lorentzian fits to the data. It can be seen that the optimization in the pump wavefront results in a 40\% stronger Raman scattering signal. 

Both the TiO$_2$ and the HAP layers are composed by a scattering powder, but since the two refractive indices are very different ($n_{TiO_2}\simeq 2.26, n_{\text{HAP}}\simeq 1.6$) most of the scattering is coming from the TiO$_2$ layer, and when optimizing the total transmission we are increasing the pump intensity in both layers. As the spontaneous Raman signal is proportional to the pump intensity~\cite{Cheng}, we expect the increase in the Raman signal collected from the HAP layer to be roughly linear with the increase in the total transmission. In Fig.~\ref{fig:Inc40}b we show the relationship between the increase in total transmission of the pump and the increase in the intensity of the collected Raman signal, for different values of the optimization. The best fit model follows a linear curve of slope 1.07$\pm$ 0.12, showing that to a very good approximation the increase in pump transmission is linear with the increase in the collected Raman intensity. 

\subsection{Estimated increase in penetration depth}
\begin{figure}[tb]
	\centering
	\includegraphics[width=1\linewidth]{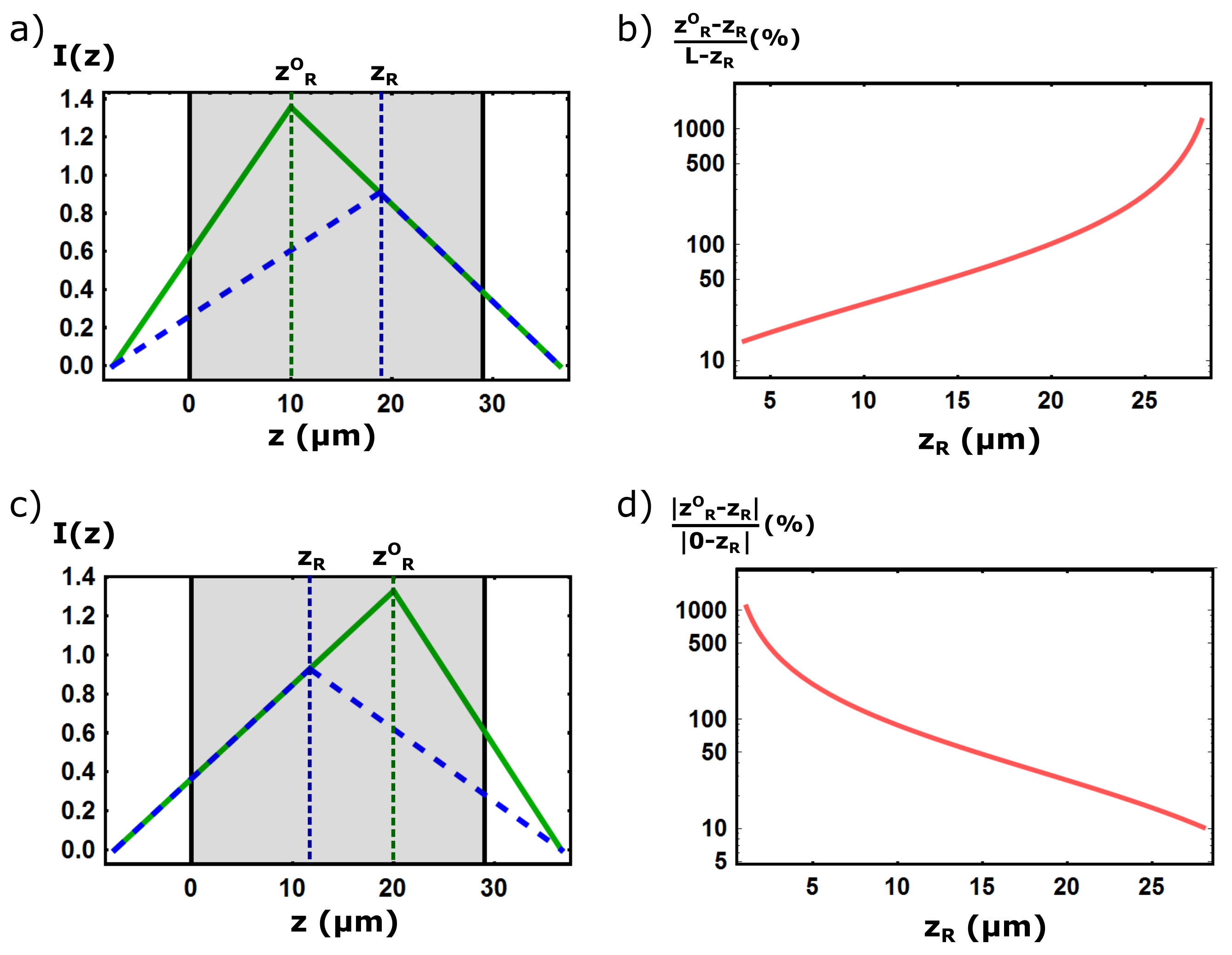}
	\caption{a) Intensity distributions of Raman elements at two different positions giving rise to the same forward emission. The dashed blue curve represents the intensity distribution of a Raman element excited with a non-optimized pump at position $z_R$. The green curve represents the intensity distribution of a Raman element at a position $z_{R}^{O}$ excited with an optimized pump (with 1.5 times higher intensity), farther away from the back-surface ($L=29\mu m$), yet resulting in the same forward emission. b) Normalized increase in the distance at which an optimized Raman element would give rise to the same forward emission as a non-optimized one, dependent on the position of the Raman element ($z_R$). Note that the vertical axis is in logarithmic scale. c) Intensity distribution of the optimized and non-optimized Raman elements resulting in the same backward emission. d) Normalized increase in the distance at which the optimized Raman element should be in order to give the same backwards emission as the non-optimized emission.}
	\label{fig:penetrationdepth}
\end{figure}

Increasing the amount of generated Raman signal means that one can detect samples from deeper into a scattering medium. To quantify how much, we assume that a Raman active medium at a depth $z_R$ produces an acceptable signal to noise ratio, and normalize to 1 the intensity of the signal produced, which will then diffuse, exit the scattering slab, and reach our detector (blue dashed curve in Fig.~\ref{fig:penetrationdepth}a,c. If wavefront shaping of the pump enhances the Raman light produced by a factor $\beta$ the active medium can be at a deeper position $z_R^0$ and still produce the same amount of measurable signal (green solid curve in Fig.~\ref{fig:penetrationdepth}a,c). 

In order to generate the same output Raman signal, according to Eq.~\ref{eq:green} and Eq.~\ref{eq:totransref} the two distributions of intensity need to have the same gradient at the edges of the sample ($z=L$ in transmission, as in Fig.~\ref{fig:penetrationdepth}a and $z=0$ in reflection, as in Fig.~\ref{fig:penetrationdepth}c. Calling $d=\left| z_{R}^{O}-z_R\right|$ we obtain
\begin{equation}
\frac{I(z_R)}{Ds_L}=\frac{\beta I(z_{R}-d)}{Ds_L +d} \, ,
\label{eqdistance}
\end{equation}
where $Ds_L=L-z_R+z_{e2}$ and $Ds_{L}^{O}=L-z_{R}^{O}+z_{e2}$.

Solving Eq.~\ref{eqdistance} for the distance $d$, we obtain that the extra distance at which a Raman element with the optimized intensity can be placed to generate the same measurable signal is given by: 
\begin{equation}
\begin{aligned}
&d(z_R)=\frac{-1}{2 \beta}\left(-\beta G(z_R)+\frac{C(z_R)}{Ds_L(z_R)}\right)\\
&+\frac{1}{2 \beta}\sqrt{\left(\beta G(z_R)-\frac{C(z_R)}{Ds_L(z_R)}\right)^2 - 4\beta (C(z_R)(1-\beta))}
\end{aligned}
\label{eq:distance}
\end{equation}
where $G(z_R)=-L+z_{e1}-z_{e2}+2z_R$ and $C(z_R)~=~(z_R+z_{e1})\cdot(L+z_{e2}-z_R)$. 

In Fig.~\ref{fig:penetrationdepth}b we plot (on logarithmic scale) the normalized increase in the distance to the output surface ($\frac{d(z_{R})}{L-z_{R}}$), achieved by the extra pump intensity as a function of $z_R$. It is interesting to note that when the Raman element is around the central region of the sample, which is, by conventional approaches, the least accessible region~\cite{vardaki2015studying}, the optimized pump is capable of achieving an extra penetration depth between \SI{30}{} and \SI{100}{}\%. Fig.~\ref{fig:penetrationdepth}d represents the equivalent case to Fig.~\ref{fig:penetrationdepth}b but when considering the Raman intensity collected in a reflection geometry, which leads to a very similar result.

\section{Discussion}

The technique presented and demonstrated in this work allows to increase the amount of pump energy density inside a scattering medium, thus increasing the amount of Raman signal produced by a sample buried deep inside it, without increasing the incident flux. As the optimization is done on the pump, instead of the Raman signal, this process can take a much shorter time than the measurement of the Raman signal itself, and can be in principle fast enough to be performed on living samples~\cite{conkey2012high}. The transmission geometry used here was chosen for simplicity; however, there is no fundamental problem that prevents to use the more common epi geometry, or even combining this technique with spatially offset Raman spectroscopy~\cite{matousek2016development,matousek2009emerging}.
The main practical limitation of this approach is that the maximum enhancement of the Raman signal that can be obtained decreases with the thickness of the sample. In fact, $M$ is proportional to the illuminated area $A \propto L^2$. Whilst the number of controlled elements $N_C$ is limited by the spatial light modulator, $M$ increases with the thickness and the total enhancement in total transmission decreases, given by Eq.~\ref{totransm}. When $M>>N_C$ Eq.~\ref{totransm} scales as approximately $1 + 2\sqrt{\sfrac{N_C}{M}}$. 
The decrease in the maximum enhancement can be counterbalanced by increasing the number of input modes, $N_C$, that is optimized.

Both the derivations and the experiments presented here assume a negligible absorption. If this is not the case it becomes even more difficult to detect a signal coming from the middle of the system if no wavefront shaping is used~\cite{vardaki2015studying,vardaki2016characterisation}, but it is nevertheless still possible to optimize the pump signal to enhance the amount of energy inside the scattering slab, given that this approach is only based on the modification of the elastically scattered light.

\section{Conclusions}
We have shown that the signal coming from a Raman active material buried deep inside a scattering medium can be enhanced by maximizing the total pump transmission via wavefront shaping. This is due to the fact that higher transmission modes store more energy inside the scattering medium than modes with low transmission~\cite{choi2011transmission}. By performing the wavefront shaping on the pump instead of the Raman signal, the process can be completed on a timescale much shorter than the Raman signal measurement itself~\cite{conkey2012high}. We experimentally shown a $\sim 48\%$ increase of the Raman signal in a well controlled and characterized system ($\sim 30 \mu$m thick TiO$_2$ layer with OD $\sim 50$ hiding a HAP sample). These results show the potential of wavefront shaping techniques for enhancing the sensitivity of detection for biomedical applications such as the measurement of breast calcifications and bone quality using deep Raman approaches. We expect that this approach can enable even deeper readout of surface enhanced Raman scattering (SERS) labeled nanoparticles and surface enhanced spatial offset Raman spectroscopy (SESORS)~\cite{stone2011surface}.

\section*{Funding Information}
A. M. Paniagua-Diaz acknowledges support from EPSRC (EP/L015331/1) through the Centre of Doctoral Training in Metamaterials (XM2). A. Ghita and N. Stone contribution was funded by EPSRC [EP/P012442/1]. J. Bertolotti acknowledges support from the Leverhulme Trust’s Philip Leverhulme Prize and the Leverhulme Trust (No. RPG-2016-129).

The research data supporting this publication are openly available from~\cite{zenodo}

\bibliography{Mendel}

\begin{thebibliography}{10}
\newcommand{\enquote}[1]{``#1''}

\bibitem{raman1928new}
C.~V. Raman and K.~S. Krishnan, \enquote{A new type of secondary radiation,}
  {\protect\JournalTitle{Nature}} \textbf{121}, 501 (1928).

\bibitem{graves1989practical}
P.~Graves and D.~Gardiner, \emph{Practical Raman Spectroscopy} (Springer,
  1989).

\bibitem{mahadevan1996raman}
A.~Mahadevan-Jansen and R.~R. Richards-Kortum, \enquote{Raman spectroscopy for
  the detection of cancers and precancers,} {\protect\JournalTitle{Journal of
  Biomedical Optics}} \textbf{1}, 31--70 (1996).

\bibitem{matousek2016development}
P.~Matousek and N.~Stone, \enquote{Development of deep subsurface raman
  spectroscopy for medical diagnosis and disease monitoring,}
  {\protect\JournalTitle{Chemical Society Reviews}} \textbf{45}, 1794--1802
  (2016).

\bibitem{matousek2009emerging}
P.~Matousek and N.~Stone, \enquote{{Emerging concepts in deep Raman
  spectroscopy of biological tissue},} {\protect\JournalTitle{Analyst}}
  \textbf{134}, 1058--1066 (2009).

\bibitem{Ghita2016}
A.~Ghita, P.~Matousek, and N.~Stone, \enquote{{Exploring the effect of laser
  excitation wavelength on signal recovery with deep tissue transmission Raman
  spectroscopy},} {\protect\JournalTitle{The Analyst}} \textbf{141}, 5738--5746
  (2016).

\bibitem{Kong2015}
K.~Kong, C.~Kendall, N.~Stone, and I.~Notingher, \enquote{{Raman spectroscopy
  for medical diagnostics — From in-vitro biofluid assays to in-vivo cancer
  detection},} {\protect\JournalTitle{Advanced Drug Delivery Reviews}}
  \textbf{89}, 121--134 (2015).

\bibitem{Akkermans2007}
E.~Akkermans and G.~Montambaux, \emph{Mesoscopic Physics of Electrons and
  Photons} (Cambridge University Press, 2007).

\bibitem{Vellekoop2007FocusingMedia}
I.~M. Vellekoop and A.~P. Mosk, \enquote{{Focusing coherent light through
  opaque strongly scattering media},} {\protect\JournalTitle{Opt. Lett.}}
  \textbf{32}, 2309 (2007).

\bibitem{mosk2012controlling}
A.~P. Mosk, A.~Lagendijk, G.~Lerosey, and M.~Fink, \enquote{Controlling waves
  in space and time for imaging and focusing in complex media,}
  {\protect\JournalTitle{Nature photonics}} \textbf{6}, 283 (2012).

\bibitem{vellekoop2008demixing}
I.~M. Vellekoop, E.~Van~Putten, A.~Lagendijk, and A.~Mosk, \enquote{Demixing
  light paths inside disordered metamaterials,} {\protect\JournalTitle{Optics
  express}} \textbf{16}, 67--80 (2008).

\bibitem{Thompson2016}
J.~V. Thompson, G.~A. Throckmorton, B.~H. Hokr, and V.~V. Yakovlev,
  \enquote{{Wavefront shaping enhanced Raman scattering in a turbid medium},}
  {\protect\JournalTitle{Optics Letters}} \textbf{41}, 1769 (2016).

\bibitem{Vellekoop}
I.~M. Vellekoop, \enquote{Feedback-based wavefront shaping,}
  {\protect\JournalTitle{Optics express}} \textbf{23}, 12189--12206 (2015).

\bibitem{conkey2012high}
D.~B. Conkey, A.~M. Caravaca-Aguirre, and R.~Piestun, \enquote{{High-speed
  scattering medium characterization with application to focusing light through
  turbid media},} {\protect\JournalTitle{Optics express}} \textbf{20},
  1733--1740 (2012).

\bibitem{Zhu1991}
J.~X. Zhu, D.~J. Pine, and D.~A. Weitz, \enquote{{Internal reflection of
  diffusive light in random media},} {\protect\JournalTitle{Physical Review A}}
  \textbf{44}, 3948--3959 (1991).

\bibitem{berkovits94}
R.~Berkovits and S.~Feng, \enquote{Correlations in coherent multiple
  scattering,} {\protect\JournalTitle{Phys. Rep.}} \textbf{238}, 135 -- 172
  (1994).

\bibitem{deboer92}
J.~F. de~Boer, M.~P. van Albada, and A.~Lagendijk, \enquote{Transmission and
  intensity correlations in wave propagation through random media,}
  {\protect\JournalTitle{Phys. Rev. B}} \textbf{45}, 658--666 (1992).

\bibitem{Albada1985}
M.~P.~V. Albada and A.~Lagendijk, \enquote{{Observation of Weak Localization of
  Light in a Random Medium},} {\protect\JournalTitle{Physical Review Letters}}
  \textbf{55}, 2692--2695 (1985).

\bibitem{Wolf1985}
P.-E. Wolf and G.~Maret, \enquote{{Weak Localization and Coherent
  Backscattering of Photons in Disordered Media},}
  {\protect\JournalTitle{Physical Review Letters}} \textbf{55}, 2696--2699
  (1985).

\bibitem{beenakker1997random}
C.~W.~J. Beenakker, \enquote{{Random-matrix theory of quantum transport},}
  {\protect\JournalTitle{Reviews of modern physics}} \textbf{69}, 731 (1997).

\bibitem{Dorokhov1984}
O.~Dorokhov, \enquote{On the coexistence of localized and extended electronic
  states in the metallic phase,} {\protect\JournalTitle{Solid state
  communications}} \textbf{51}, 381--384 (1984).

\bibitem{choi2011transmission}
W.~Choi, A.~P. Mosk, Q.-H. Park, and W.~Choi, \enquote{{Transmission
  eigenchannels in a disordered medium},} {\protect\JournalTitle{Physical
  Review B}} \textbf{83}, 134207 (2011).

\bibitem{Davy2015}
M.~Davy, Z.~Shi, J.~Park, C.~Tian, and A.~Z. Genack, \enquote{Universal
  structure of transmission eigenchannels inside opaque media,}
  {\protect\JournalTitle{Nature communications}} \textbf{6}, 6893 (2015).

\bibitem{koirala2017inverse}
M.~Koirala, R.~Sarma, H.~Cao, and A.~Yamilov, \enquote{Inverse design of
  perfectly transmitting eigenchannels in scattering media,}
  {\protect\JournalTitle{Phys. Rev. B}} \textbf{96}, 054209 (2017).

\bibitem{vardaki2015studying}
M.~Z. Vardaki, B.~Gardner, N.~Stone, and P.~Matousek, \enquote{Studying the
  distribution of deep raman spectroscopy signals using liquid tissue phantoms
  with varying optical properties,} {\protect\JournalTitle{Analyst}}
  \textbf{140}, 5112--5119 (2015).

\bibitem{fick1855}
A.~Fick, \enquote{Uber diffusion,} {\protect\JournalTitle{Poggendorff's Annalen
  der Physik und Cheimie}} \textbf{94}, 59 (1855).

\bibitem{Ojambati2016CouplingMedium}
O.~S. Ojambati, H.~Yilmaz, A.~Lagendijk, A.~P. Mosk, and W.~L. Vos,
  \enquote{{Coupling of energy into the fundamental diffusion mode of a complex
  nanophotonic medium},} {\protect\JournalTitle{New Journal of Physics}}
  \textbf{18} (2016).

\bibitem{kim2012maximal}
M.~Kim, Y.~Choi, C.~Yoon, W.~Choi, J.~Kim, Q.-H. Park, and W.~Choi,
  \enquote{{Maximal energy transport through disordered media with the
  implementation of transmission eigenchannels},} {\protect\JournalTitle{Nature
  photonics}} \textbf{6}, 581--585 (2012).

\bibitem{choi2015preferential}
W.~Choi, M.~Kim, D.~Kim, C.~Yoon, C.~Fang-Yen, Q.-H. Park, and W.~Choi,
  \enquote{{Preferential coupling of an incident wave to reflection
  eigenchannels of disordered media},} {\protect\JournalTitle{Scientific
  reports}} \textbf{5} (2015).

\bibitem{Vellekoop2008UniversalMaterials}
I.~M. Vellekoop and A.~P. Mosk, \enquote{{Universal Optimal Transmission of
  Light Through Disordered Materials},} {\protect\JournalTitle{Physical Review
  Letters}} \textbf{101}, 120601 (2008).

\bibitem{goetschy2013filtering}
A.~Goetschy and A.~Stone, \enquote{Filtering random matrices: the effect of
  incomplete channel control in multiple scattering,}
  {\protect\JournalTitle{Physical review letters}} \textbf{111}, 063901 (2013).

\bibitem{Marcenko1967}
V.~A. Marcenko and L.~A. Pastur, \enquote{{Distribution of eigenvalues for some
  sets of random matrices},} {\protect\JournalTitle{Mathematics of the
  USSR-Sbornik}} \textbf{1}, 457--483 (1967).

\bibitem{Jacques2013}
S.~L. Jacques, \enquote{{Optical properties of biological tissues: a review},}
  {\protect\JournalTitle{Physics in Medicine and Biology}} \textbf{58},
  R37--R61 (2013).

\bibitem{vo2014biomedical}
T.~Vo-Dinh, \emph{{Biomedical Photonics Handbook: Fundamentals, Devices and
  Techniques}}, vol.~1 (CRC press, 2014).

\bibitem{Cao2015openchannelsabs}
R.~Sarma, A.~Yamilov, S.~F. Liew, M.~Guy, and H.~Cao, \enquote{{Control of
  mesoscopic transport by modifying transmission channels in opaque media},}
  {\protect\JournalTitle{Physical Review B}} \textbf{92}, 214206 (2015).

\bibitem{mucalo2015hydroxyapatite}
M.~Mucalo, \emph{Hydroxyapatite (HAp) for biomedical applications} (Elsevier,
  2015).

\bibitem{Cheng}
J.~Cheng and X.~S. Xie, \emph{Coherent Raman scattering microscopy} (CRC press,
  2016).

\bibitem{vardaki2016characterisation}
M.~Z. Vardaki, P.~Matousek, and N.~Stone, \enquote{Characterisation of signal
  enhancements achieved when utilizing a photon diode in deep raman
  spectroscopy of tissue,} {\protect\JournalTitle{Biomedical optics express}}
  \textbf{7}, 2130--2141 (2016).

\bibitem{stone2011surface}
N.~Stone, M.~Kerssens, G.~R. Lloyd, K.~Faulds, D.~Graham, and P.~Matousek,
  \enquote{Surface enhanced spatially offset raman spectroscopic (sesors)
  imaging--the next dimension,} {\protect\JournalTitle{Chemical Science}}
  \textbf{2}, 776--780 (2011).

\bibitem{zenodo}
\url{http://doi.org/10.5281/zenodo.1319374}.

\end{thebibliography}

\bibliographyfullrefs{Mendel}

\end{document}